# Designing Gamified Social Interaction for Gen Z in the Metaverse: A Framework-Oriented Systematic Literature Review

Kris (Baitong) XIE
*School of Information and Technology*
*Murdoch University*
*Murdoch, Western Australia, Australia*
*35352611@student.murdoch.edu.au*
*0009-0004-3798-7299*
*(Corresponding Author)*

Mohd Fairuz SHIRATUDDIN
*School of Information and Technology*
*Murdoch University*
*Murdoch, Western Australia, Australia*
*f.shiratuddin@murdoch.edu.au*
*0000-0002-9529-6320*

Mostafa HAMADI
*School of Information and Technology*
*Murdoch University Murdoch,*
*Western Australia, Australia*
*CCIT, University of Doha for Science*
*and Technology*
*Doha, Qatar*
*m.hamadi@murdoch.edu.au*

Joo Yeon PARK
*School of Information and Technology*
*Murdoch University*
*Murdoch, Western Australia, Australia*
*jooyeon.park@murdoch.edu.au*
*0000-0002-5231-5405*

Thao DUONG
*School of Information and Technology*
*Murdoch University*
*Murdoch, Western Australia, Australia*
*thao.duong@murdoch.edu.au*
*0000-0003-2294-3619*

*Abstract*— Gamification plays a pivotal role in enhancing user engagement in the Metaverse, particularly among Generation Z users who value autonomy, immersion, and identity expression. However, current research lacks a cohesive framework tailored to designing gamified social experiences in immersive virtual environments. This study presents a framework-oriented systematic literature review, guided by PRISMA 2020 and SPIDER, to investigate how gamification is applied in the Metaverse and how it aligns with the behavioral needs of Gen Z. From 792 screened studies, seventeen high-quality papers were synthesized to identify core gamification mechanics, including avatars, XR affordances, and identity-driven engagement strategies. Building on these insights, we propose the Affordance-Driven Gamification Framework (ADGF), a conceptual model for designing socially immersive experiences, along with a five-step design process to support its real-world application. Our contributions include a critical synthesis of existing strategies, Gen Z-specific design considerations, and a dual-framework approach to guide researchers and practitioners in developing emotionally engaging and socially dynamic Metaverse experiences.

## I. Introduction

The Metaverse, first conceptualized by Stephenson in Snow Crash in 1992 [1], has evolved into a persistent, multi-user ecosystem that blends digital and physical experiences through eXtended Reality (XR) modalities: Augmented Reality (AR), Mixed Reality (MR), and Virtual Reality (VR) [2, 3]. Enabled by infrastructures such as AI, blockchain, and spatial computing, it offers a socio-technical space for social, cultural, and behavioral innovation.

Gamification, the strategic use of game design elements in non-game contexts, has proven effective in enhancing motivation, engagement, and behavioral outcomes across education [4], healthcare [5], workplace productivity [6], and social networking [7]. For Generation Z, digital natives who value autonomy, self-expression, and interactive experiences, gamification can sustain engagement and deepen emotional connection [8, 9]. In immersive Metaverse environments, it has been shown to enhance presence, reduce cognitive fatigue, and support positive psychological states [10-12]. Despite its promise, current gamification frameworks for the Metaverse are fragmented and often lack adaptability to its interdisciplinary and dynamic nature. The variability of XR modalities demands alignment between gamification mechanics, platform affordances, and user psychology. Few studies comprehensively map gamification affordances to immersive environments tailored for Gen Z.

To address these gaps, this study conducts a Systematic Literature Review (SLR) guided by PRISMA 2020 [13] and SPIDER [14], synthesizing key gamification mechanics, theoretical models, and their alignment with Gen Z behaviors. Building on the Mechanics-Dynamics-Aesthetics (MDA) model and affordance theory, we propose the Affordance-Driven Gamification Framework (ADGF), a five-stage conceptual model for designing and evaluating socially engaging gamified Metaverse experiences.

This study is guided by three research questions: RQ1: What gamification mechanics and design strategies are commonly considered to be used in the Metaverse currently? RQ2: What conceptual or theoretical frameworks have been proposed for gamification in Metaverse environments? RQ3: Which gamification mechanics potentially align with Generation Z users' behaviors, needs, and motivations in the Metaverse? By answering these questions, the study aims to (1) review existing gamification frameworks and evaluate their applicability in immersive contexts, and (2) develop the ADGF to address the social, perceptual, and technological needs of gamified Metaverse environments for Gen Z.

## II. BACKGROUND

### A. The Metaverse

The Metaverse is broadly defined as a convergent socio-technical ecosystem integrating VR, AR, MR, AI, blockchain, IoT, and next-generation communications (5G/6G) [3, 15], [16]. Core features include immersion, advanced computing, socialization, and decentralization [3]. It blurs boundaries between online and offline identities, especially for Generation Z, whose digital selves often extend their real-world personas [2]. The study conceptualizes the Metaverse as a hyper-spatio-temporal, socially embedded virtual world, characterized by multi-technology convergence (e.g., IoT, digital twins, blockchain), sociality through avatars in immersive, networked spaces, and hyper-spatial temporality, which transcends physical time and space [17]. Two main perspectives dominate: the infrastructural vision, focusing on technological architecture, and the performative vision, emphasizing user agency in shaping meaning, identity, and social presence. This study adopts a performative stance while recognizing that social experiences are technologically scaffolded by affordances such as spatial computing, persistent avatars, and multimodal sensory input.

### B. Maintaining the Integrity of the Specifications

Gen Z (born 1997–2009) are digital natives with the highest digital skills of any generation [18, 19]. They have never experienced life without the Internet [20, 21] and prefer mobile apps, social media, and video content over traditional modes of communication. Research highlights their preference for instant gratification, self-expression, and experimentation with emerging technologies [8, 9]. These attributes make them active participants and co-creators in immersive Metaverse environments, responding positively to personalized, gamified experiences.

### C. Gamification

Gamification is the application of game design elements in non-game contexts [22], distinguished by its structured, goal-oriented, and user-centered design [23]. Effective gamification frameworks draw on motivational theory, usability, and behavioral outcomes [24], incorporating mechanics such as feedback, personalization, and progression. Traditional gamification often uses points, badges, and leaderboards. However, in immersive environments, these mechanics require reinterpretation through spatial, sensory, and identity-based affordances. Identity expression shifts from static profiles to embodied avatars, gesture-based interaction, and co-presence. In the Metaverse, gamification has the potential to enhance persistent social presence, identity construction, and co-creation. Yet, most existing frameworks, developed for conventional domains, lack the flexibility required for the open-ended, interdisciplinary nature of immersive environments. Strategic alignment between mechanics, XR affordances, and user psychology is essential, particularly for Gen Z users who expect seamless integration of technology, personalization, and social engagement.

## III. METHODOLOGY

### A. Research Design

This study employs a Systematic Literature Review (SLR) to identify, evaluate, and synthesize research on gamification in the Metaverse, with particular attention to Generation Z. The process follows PRISMA 2020 for procedural rigor and transparency [13], complemented by the SPIDER framework [14] to refine search logic and ensure thematic relevance. PRISMA ensured systematic identification, screening, and selection of articles. SPIDER guided keyword development and inclusion parameters across five dimensions: Sample, Phenomenon of Interest, Design, Evaluation, and Research Type. SPIDER was particularly useful in managing the interdisciplinary nature of the topic and in focusing on Gen Z, gamification, and immersive XR environments.

### B. Search Strategy, Inclusion and Exclusion Criteria

Five major databases were searched for peer-reviewed publications between January 1, 2020, and February 1, 2025 (Table I). The Search terms were informed by SPIDER and refined through Boolean operators to capture three strategies: (1) Core Context: Gamification + Metaverse, (2) Outcomes/Effects: Gamification + engagement/motivation/immersion/social interaction, and (3) Framework Focus: Gamification + conceptual framework/theoretical model. The initial search yielded 792 results (Table II). Table III shows the inclusion and exclusion criteria.

Table I.    DIGITAL LIBRARIES

| Digital Library | URL |
|---|---|
| IEEE Xplore | http://ieeexplore.ieee.org |
| Scopus | http://www.scopus.com |
| Web of Science | https://www.webofscience.com |
| Taylor & Francis Online | https://www.tandfonline.com |
| SAGE | https://journals.sagepub.com |

Table II.    SEARCH TERMS & RESULTS

| Strategy | Search terms used in databases and search engines | Results (per DB) |
|---|---|---|
| 1 | ("gamification" OR "game mechanics" OR "game-based interaction") AND ("Metaverse") | Scopus N=142<br>IEEE Xplore N=47<br>WOS N=91<br>Sage N=50<br>Taylor & Francis Online N=159 |
| 2 | ("gamification" OR "game mechanics") AND ("user engagement" OR "motivation" OR "immersion" OR "social interaction") AND ("Metaverse") | Scopus N=37<br>IEEE Xplore N=16<br>WOS N=25<br>Sage N=33<br>Taylor & Francis Online N=126 |
| 3 | ("gamification" OR "game mechanics") AND ("conceptual framework" OR "theoretical model") AND ("Metaverse") | Scopus N=4<br>IEEE Xplore N=2<br>WOS N=2<br>Sage N=13<br>Taylor & Francis Online N=45 |
| | | Total N=792 |

| Table III. | CRITERION | |
|---|---|---|
| **Criterion** | **Inclusion** | **Rationale** |
| Period | 2020-2025 | Focus on current XR/Metaverse trends |
| Type | Peer-reviewed journals & conference papers | Ensure quality |
| Language | English | Maintain consistency |
| Topic Relevance | Metaverse, Gamification, Social Interaction | Align with RQs |
| Domain | Education, cultural heritage, marketing, entertainment | Transferable gamification insights |

*C. Screening Process*

The PRISMA process (Figure I) reduced the dataset from 792 to 17 high-quality studies after: (1) removing duplicates (n=394); (2) excluding non-peer-reviewed, non-English, pre-2020 studies (n=37); (3) filtering for relevance to Metaverse + gamification + social interaction (n=431 excluded), and (4) applying domain/context criteria (n=309 excluded).

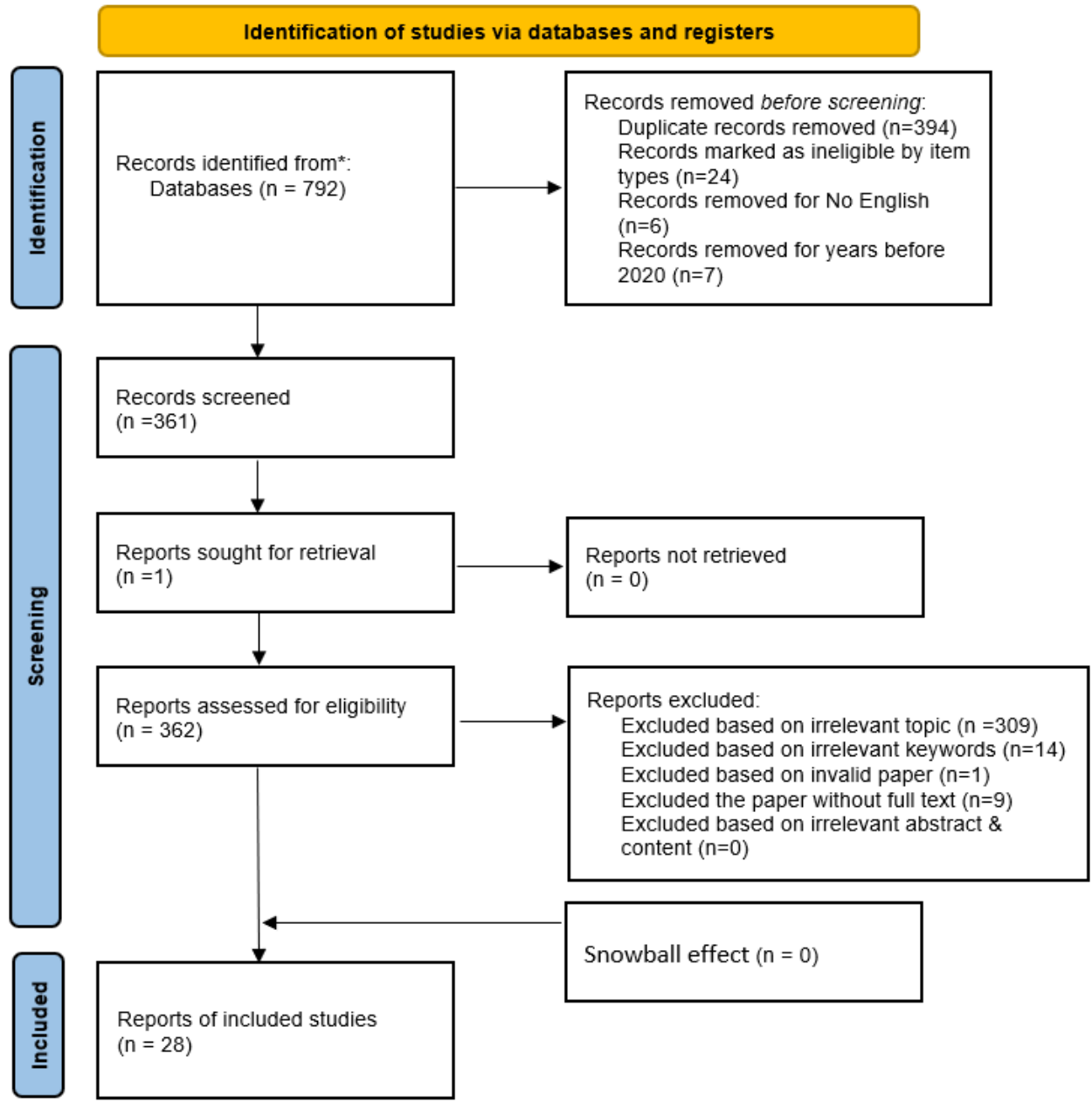

FIGURE I. PRISMA FLOW DIAGRAM

*D. Eligibility Assessment*

The SPIDER framework was reapplied for detailed eligibility, focusing on: (1) Sample relevance to Gen Z or a comparable demographic; (2) Gamification Depth beyond surface mentions; (3) Framework Presence or theoretically informed design, and (4) Methodological Transparency for replicability. Only studies meeting these criteria were retained. Those not explicitly targeting Gen Z but with age-aligned participants (e.g., [11, 25]) were included.

The methodological rigor and relevance of the 17 included studies were independently evaluated by all three authors using a predefined SPIDER-based appraisal matrix encompassing sample definition, research design transparency, and clarity of evaluation. Each study was reviewed separately, and evaluators recorded their judgments in a shared coding template to ensure consistent application of the criteria. Discrepancies were discussed through iterative consensus meetings until convergence was achieved, ensuring that interpretations reflected collective agreement rather than individual bias. This intersubjective assessment procedure prioritized transparency, reflexivity, and reproducibility, serving the same quality-assurance function as formal scoring tools while remaining philosophically consistent with the qualitative and conceptual orientation of the synthesis. Although structured instruments such as MMAT, CASP, or JBI were considered, they were deemed unsuitable given the conceptual heterogeneity of the corpus; hence, the adopted intersubjective protocol was selected as a methodologically sound and contextually appropriate alternative. This interpretive consensus process aligns with qualitative evidence synthesis traditions, where intersubjective validation is used to ensure methodological transparency and shared interpretive coherence among reviewers.

### E. Data Extraction

Using SPIDER, each retained study was coded for domain, sample, phenomenon, and key insights related to gamification (Table IV).

Table IV. SPIDER DATA EXTRACTION - A SUMMARY OF THE INCLUDED STUDIES

| Paper | Domain | S (Sample) | PI (Phenomenon) | Key Gamification Insights |
|---|---|---|---|---|
| 1. [26] | Entertainment/Tourism / Metaverse Experience Design | 671 South Korean participants (59% female, mostly aged 21–30). All had experience with Metaverse concerts (e.g., Justin Bieber's performance). Includes Gen Z representation (Occupied 59% in the survey). | Investigates the gamification affordances of achievement, identity, competition, and self-expression. Link these to perceived value (effectiveness, social gain, enjoyment), flow state, and engagement. | - Identity and competition affordances strongly influence social gain and enjoyment.<br>- Flow state is a key mediator between gamification and engagement.<br>- Gender moderates the effect of enjoyment on flow-males show stronger flow from enjoyment. |
| 2. [12] | Design / Gamification / Metaverse / Positive Psychology / User Experience | 17 gamification designers from 14 countries (e.g., USA, India, Sweden), working in MNCs and gamification firms. | Explores gameful experience from the perspective of designers in the Metaverse era. | Designers aim to create gameful experiences that trigger interest, reward behavior, elicit curiosity, and immerse users in the Metaverse. XR technologies and psychological engagement strategies increasingly shape gamification. |
| 3. [11] | Tourism/Event Management / Consumer Behavior in XR | Mixed-methods study with 3 focus groups (20 participants) and a quantitative survey (219 participants, average age 31.7, 51% female). | Investigates how gamification mitigates negative affect and enhances authenticity and behavioral intention in metaverse cultural events. | - Gamification reduces user confusion and improves navigation in complex metaverse environments.<br>- It helps users focus their attention and reduces negative affect (e.g., frustration, anxiety).<br>- Gamification enhances users' ability to imagine cultural events and perceive authenticity, which is key to behavioral intentions (e.g., visiting a real event).<br>- Proposed gamification mechanics include challenges, quizzes, avatar customization, and scavenger hunts.<br>The study emphasizes the moderating role of gamification, which mitigates the negative impacts of low attention in XR environments. |
| 4. [27] | Educational Technology / Human–Computer Interaction / UX Research | 690 valid survey responses from Generation Z students in China. Age range: 18–30 (majority aged 25–30). Gender: 48% male, 52% female. Education level: High school to postgraduate. All participants had prior experience using gamified learning platforms. | Investigates how self-regulation and Metaverse features (social interaction, virtual avatars, immersion) influence Gen Z's attitudes and intentions to adopt gamified learning platforms. | - Gamified platforms with Metaverse features foster immersive, personalized, and socially engaging learning environments.<br>- Self-regulation mechanisms (goal setting, progress tracking, distraction control) enhance enjoyment and motivation.<br>- Gen Z learners value social connectivity, avatar customization, and immersive experiences in gamified learning. |
| 5. [28] | Language Education / Metaverse Pedagogy / Educational Technology | 55 undergraduate students (ages 19–23, mean age 20.9; 23 males, 32 females) from Hong Kong Polytechnic University. | Application of gamified constructivist teaching in Metaverse-based language learning using the Questverse platform. | - Gamification supports immersive, autonomous learning through roleplay, missions, and feedback systems.<br>- Students responded positively to personalized avatars and interactive tasks. |
| 6. [25] | Museum Studies / Cultural Heritage / Gamified Exhibition Design | 110 university students (70% female, aged 20s) from Korea with high familiarity with the metaverse. Five expert evaluators from the game development field also participated. | Investigates how gamification enhances engagement, emotional resonance, and sustainability in a metaverse exhibition linked to an offline museum. | - Gameful experiences (nudge, flow, alternate reality, hedonic) were effectively implemented.<br>- Missions and symbolic acts (e.g., burning collected items) fostered emotional healing and engagement. |
| 7. [29] | Conceptual Technology Review – Virtual Reality | Conceptual paper; no empirical sample. | Explores gamification mechanics (avatars, badges, guilds, digital goods, extended reality, AI/ML) within the metaverse and their implications for user engagement. | - Gamification in the metaverse is driven by personalization (avatars), social bonding (guilds), and immersive tech (XR, AI).<br>- Emphasizes the role of gamification in bridging social gaps, especially post-COVID. |
| 8. [30] | Tourism Marketing / Gamified | 94 users (ages 8–45) and 9 tour agency representatives in Thailand. | Evaluates how gamification in Roblox-based virtual tourism (Anytime Tour) enhances user engagement and supports marketing for tour agencies. | - Combines storytelling, quests, challenges, rewards, and social interaction to create immersive tourism experiences.<br>- Gamification drives real-world travel interest and brand engagement. |
| 9. [10] | Conceptual Framework / Game Design / Digital Culture | Conceptual and theoretical paper; no empirical sample. | Explores the integration of gamification principles into metaverse environments, emphasizing user motivation, engagement, and behavioral change. | - Gamification in the metaverse fosters flow, motivation, and emotional engagement.<br>- The paper categorizes game elements into dynamics, mechanics, and components, and links them to player types and entertainment styles (hard fun, easy fun, serious fun, social fun). |
| 10. [31] | XR & Telecom | An experimental study with 10 participants (aged 25–32), both online (Amazon MTurk) and offline. | Explores gamified experiences across eight XR-based Multiverse realms using VR/AR headsets and social robots. | - Gamified XR experiences vary in effectiveness across realms.<br>- Combining realms (e.g., AR + haptics) can optimize engagement and reduce failure rates. |
| 11. [32] | Application with AR and VR for Cultural Heritage Preservation in the Metaverse | Mixed methods:<br>10 qualitative participants (students, educators, curators)<br>77 survey respondents (94.8% aged 18–30) | Investigate how gamification and immersive technologies (such as AR/VR) in the metaverse can enhance cultural heritage education and engagement. | - Gamification (points, challenges, rewards) combined with AR/VR fosters emotional connection and learning.<br>- Tailored personas (student, researcher, tourist) guide design. |
| 12. [33] | Educational Design / Gamification Theory / Metaverse Pedagogy | The study focuses on Gen Z learners and their affinity for gameful experiences in virtual worlds | Identifies five metaverse world types (Zepeto, Roblox, Gather. town, Fortnite) that deliver gameful experiences to support sustainable learning. | Gameful experiences in metaverse worlds (e.g., quests, competition, exploration) enhance motivation and engagement.<br>The Escape Room world offers the richest experiential blend. |
| 13. [34] | Language Education / Reading Engagement / Metaverse Pedagogy | 93 fifth-grade students in Taiwan (46 experimental, 47 control). 8-week reading program using Gather. Town. | Investigate how a gamified metaverse space affects reading interest and flow perception. | Gamified metaverse fosters exploratory learning and personalized engagement.<br>NPCs and leader board mechanics enhance motivation and social presence. |
| 14. [35] | Metaverse Learning / Motivation Psychology / Gamification Design | - Study 1: 177 participants (Roblox-based learning scenario)<br>- Study 2: 400 participants (Roblox Bayside High School) | Examines how gamification mechanics affect learning motivation and satisfaction in metaverse learning. | Reward and feedback are most effective in enhancing motivation and satisfaction.<br>Gamification is most effective when it supports both intrinsic and extrinsic motivation. |
| 15. [36] | Library Science / Educational Technology / Gamified Orientation | 80 first-year university students in Thailand (40 experimental, 40 control).<br>A gamified library program using a VR-based digital twin of the library built in Unity. | Examines how a gamified metaverse library program affects students' knowledge acquisition and their anxiety about using the library. | - Gamified metaverse platforms enhance engagement and motivation, but real human interaction is crucial for reducing anxiety.<br>- NPCs and immersive design support exploratory learning, but emotional connection is limited. |
| 16. [37] | Gamified Learning / Immersive Education / AR/VR in Higher Ed | University-level mathematics students in Lithuania. Specific demographics are not detailed, but the context involves Gen Z learners in STEM courses. | Design and implementation of AR-based escape rooms using the Metaverse app (Studio Gometa) to teach differential equations and calculus. Focus on gamification mechanics, such as puzzles, feedback loops, and immersive narratives. | Gamified AR escape rooms in the Metaverse foster active learning, problem-solving, and personalized learning paths.<br>Narrative-driven design and feedback loops enhance immersion and mastery. |
| 17. [38] | Marketing / XR / Gamification / Metaverse / Consumer behavior | 688 participants from India and Morocco (Gen Z and Millennials), all with prior Metaverse experience.<br>Used the Spatial platform for immersive exposure before the survey. | Investigates how XR-based gamification marketing activities (entertainment, interaction, trendiness, intimacy, novelty) influence brand equity, brand love, and engagement in Metaverse environments. | - XR-based gamification in the Metaverse enhances consumer-brand relationships, especially through customization, emotional storytelling, and immersive interaction.<br>- Brand authenticity and avatar personalization are key to building trust and loyalty. |

## IV. RESULT AND DISCUSSION

After the text edit has been completed, the paper is ready for the template.

*A. Description of Studies*

*1) Type of Research & Methodology approaches*

The selected studies span quantitative experiments and surveys (e.g., structural equation modelling) [26-28, 38], qualitative interviews and case studies [12, 25, 33], mixed methods [11, 32, 36], conceptual/theoretical work [10, 29], quasi-experimental classroom evaluations [34], and XR-based experiments [31, 37]. This methodological breadth underscores the field's interdisciplinarity and reinforces the need for design frameworks that accommodate multiple inquiry modes when targeting Gen Z experiences in immersive contexts.

*B. Thematic findings*

*1) Identified Gamification Mechanics in the Metaverse*

*a) The role of XR technologies in gamification design in the Metaverse*

XR functions as a foundational enabler rather than a passive visual layer, structuring spatial, social, and psychological dimensions of interaction to support immersion, affect, and presence [11, 25, 26]. In education, XR augments story- and challenge-based mechanics [32, 37], while conceptual work positions XR as a core layer for spatial storytelling and social bonding [10, 29]. Commercially, XR deepens brand attachment via interactive storytelling and avatar-based personalization [38].

*b) Avatar*

Avatars, which are a core gamification element [23], drive identity construction, self-expression, and social presence; key levers for flow [39] and sustained motivation among Gen Z [9, 10, 26-30]. Beyond enjoyment, avatars signal recognition and status, shaping social dynamics and symbolic value in virtual spaces [26]. Avatars function as multidimensional affordances in the gamified Metaverse, offering psychological, social, and emotional benefits that are particularly resonant for Gen Z users. Their role is central to the expression of identity, sustained motivation, and the development of meaningful interactions within immersive environments.

*c) Other gamification mechanics*

Leaderboards, badges, and digital goods support progression and virtual ownership [10, 29, 30, 34, 36], while mission-based structures, NPC guidance, and feedback sustain attention and repeat participation [11, 25, 28, 30]. Team/guild systems, progress tracking, tokens, quests, and narrative interactions scaffold socially embedded, goal-oriented experiences, though systematic validation in Metaverse-specific contexts remains limited. While avatars remain central for identity expression and co-presence, the collective use of leaderboards, badges, missions, guilds, and reward structures provides essential scaffolding for immersive, gamified experiences. However, there remains a lack of research that systematically validates how these mechanics are designed, structured, and operationalized in Metaverse-specific contexts. Their applicability, scalability, and effectiveness, especially when mediated by XR technologies, require further empirical research.

*2) Applied gamification frameworks/theories*

To address Research Question 2, this section critically examines the theoretical foundations and gamification frameworks employed in the selected literature. As gamification becomes an increasingly central strategy for designing immersive and engaging experiences in Metaverse environments, particularly across domains such as education, marketing, and cultural heritage, researchers have adopted a variety of structured models to guide its implementation.

The frameworks and theories analyzed here are not ad hoc proposals. They have either undergone empirical validation or have been repeatedly cited across the reviewed studies, reflecting their scientific rigor and disciplinary credibility. This ensures that the discussion focuses on academically grounded models with demonstrated relevance to user motivation, participation, and social interaction in gamified virtual spaces.

By synthesizing how these frameworks shape the structure and delivery of gamified Metaverse experiences, this section lays the conceptual groundwork for understanding effective design strategies tailored to Generation Z users. In doing so, it provides a theoretical lens through which key affordances and interaction mechanisms can be evaluated and operationalized within the broader context of immersive gamification.

The review converges on academically grounded models, i.e. MDA [40], Flow Theory [39], Gameful Experience Framework [12], GCTM [28], gamification affordances [26, 41-43], and Self-Determination Theory [44]; as recurrent lenses for structuring and evaluating immersive gamified experiences.

*a) MDA framework*

The MDA framework (Figure II) clarifies how mechanics produce dynamics and aesthetic outcomes. In the Metaverse Anytime Tour tourism Roblox-based game [30], quests/points/challenges foster cooperation and repeat engagement, translating into immersive storytelling and real-world behavioral intentions [10, 40].

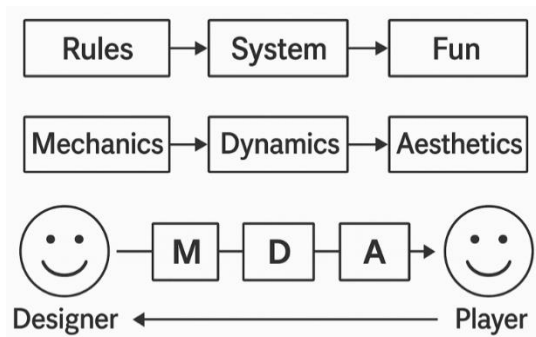

FIGURE II. THE MDA FRAMEWORK ADAPTED FROM [40]

*b) Flow theory*

Identity/competition/self-expression affordances elevate perceived value (effectiveness, social gain, enjoyment), catalyzing flow and engagement [26]. Design features such as progressive challenge, clear goals, feedback, and safe experimentation consistently elicit a state of flow [10, 12, 25].

*c) Gameful Experience Framework*

Thomas' four categories [12], i.e. nudge, flow, alternate reality, and hedonic, map directly to design/evaluation criteria (Figure II). Applied in the study [25], missions and interactive storytelling supported cognitive, emotional, and behavioral engagement with sensitive themes.

*d) GCTM framework*

GCTM integrates constructivist phases, i.e., Game World, Game Rule, Roleplay, Mission, and Evaluate, operationalized in Questverse via avatars, voice scoring, chatbots, and badges. Evaluation is conducted using the 5I model (Immersion, Insight, Imagination, Interaction, Intelligence) to enhance motivation and reduce anxiety [28].

*e) Gamification Affordances in the Metaverse*

Key affordances include achievement, identity, competition, self-expression, autonomy, interactivity, rewards, feedback, cooperation, and narrative [26, 41-43]. These drive perceived value and flow, but mappings remain fragmented and context-dependent.

*f) Self-Determination Theory*

Autonomy, competence, and relatedness are activated via self-directed paths, progressive challenges/feedback, and social/NPC interactions; aligning mechanics with SDT improves sustained participation, learning outcomes, and virtual presence [34-36, 44].

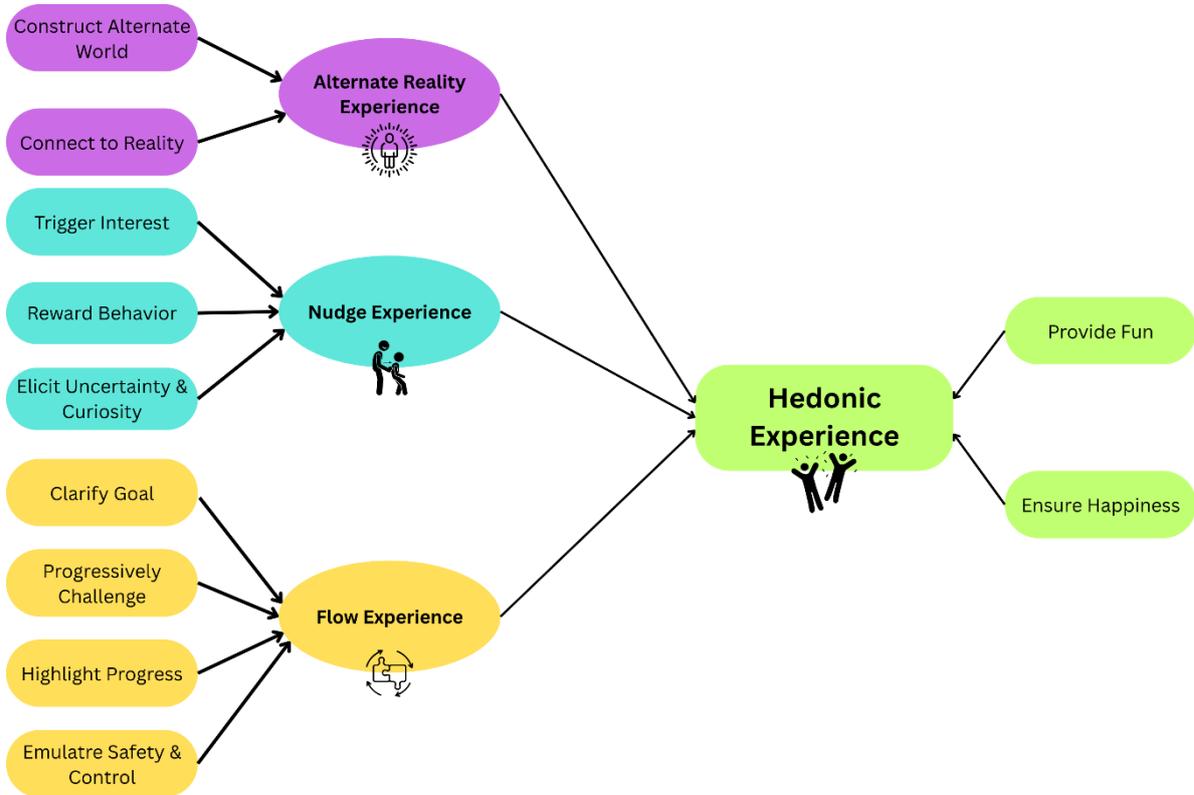

FIGURE III. GAMEFUL EXPERIENCE FRAMEWORK ADAPTED FROM [12]

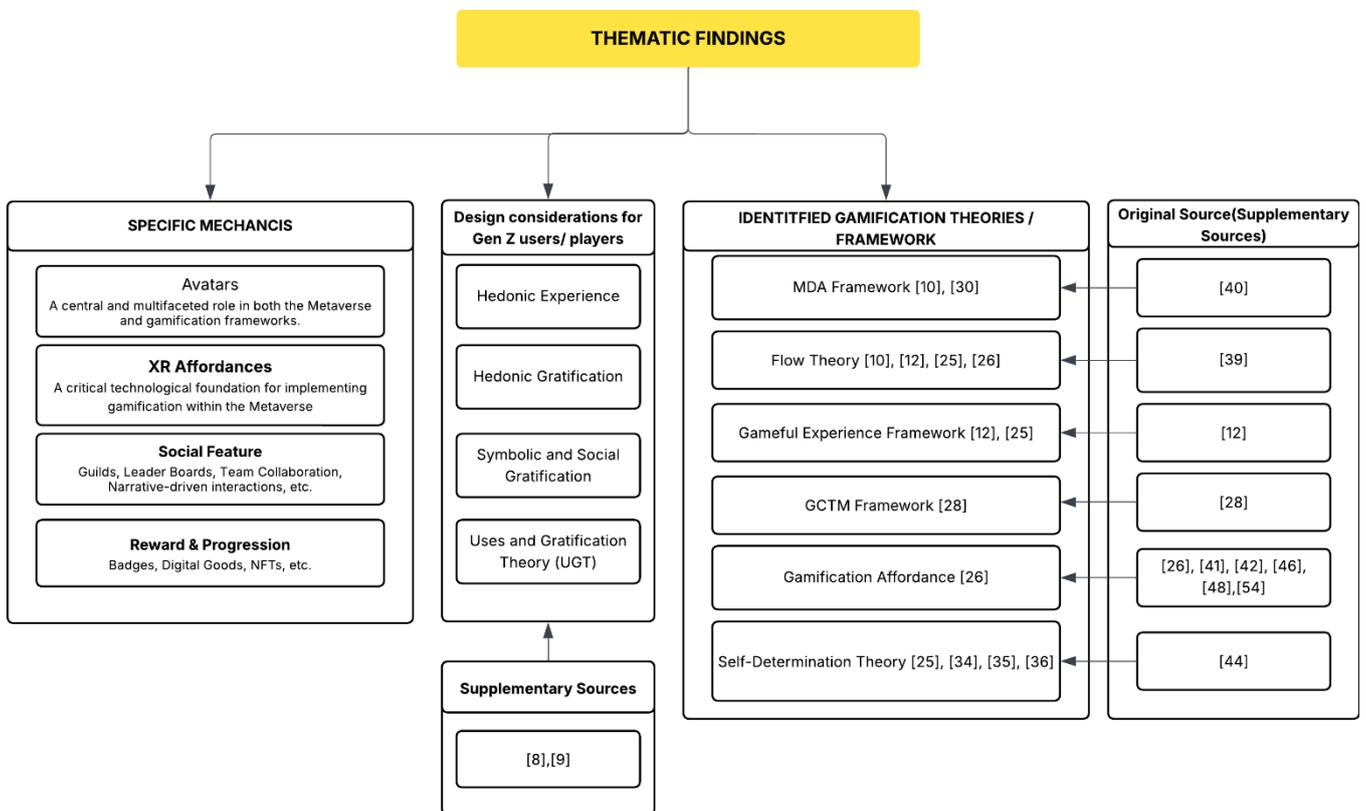

FIGURE IIII. THEMATIC FINDINGS

*3) Design considerations for Gen Z users/players*

*a) Hedonic Experience & Gratification*

Hedonic enjoyment and emotional gratification, not just extrinsic rewards, are central for Gen Z [8, 9, 12, 45]. Avatar customization, competitive/co-operative play, and social validation meet symbolic and social needs, while progression mechanics sustain flow and presence [26].

*4) Thematic Findings*

Avatars, XR-enabled affordances, and socially oriented mechanics are prevalent in Gen Z-relevant designs. Yet a unified framework linking user psychology, platform affordances, and mechanics remains underdeveloped, motivating the model proposed next. Figure IV confirms the thematic findings, illustrating the prominence of avatars, XR-based affordances, and socially oriented game mechanics in shaping gamified experiences for Generation Z within the Metaverse. Despite the richness of existing insights, the analysis reveals a lack of unified theoretical frameworks that can systematically inform design practices. Building on this foundation, we propose the Affordance-Driven Gamification Framework (ADGF). This conceptual model integrates XR-enabled, identity-based, and avatar-driven affordances to guide the design of socially immersive and emotionally engaging gamified experiences tailored to Generation Z within the Metaverse.

*C. Investigating the Gamification Framework in the Metaverse*

Existing frameworks emphasize psychological outcomes (flow, motivation, engagement) [12, 28, 39, 40, 44], but often start from preselected mechanics rather than user-centered interpretations of affordances and goals. We therefore reorient the design to begin with users' psychological/behavioral states, then map affordances, and finally select mechanics.

*1) Affordance-Driven Gamification Framework (ADGF)*

Grounded in affordance theory [46, 47] and gamification research on actionable properties [48-52], ADGF integrates the affordance→psychological→behavioral (Figure V) chain [42, 43, 50, 51] with a reverse-feedback emphasis PGG model where users' intrinsic needs and behaviors shape perceived affordances, which then guide mechanics (Figure VI). Identity construction/self-expression affordances (e.g., avatar customization, persistent identifiers) directly support autonomy and recognition and can be operationalized by specific mechanics [26, 53-55]. This psychology-first approach is particularly suited to co-creative, socially persistent Metaverse contexts.

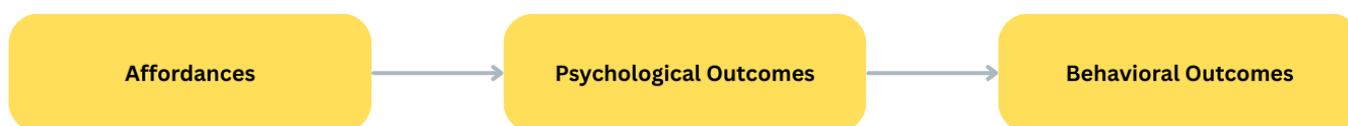

FIGURE V. THE PERSPECTIVE OF GAMIFICATION AFFORDANCES ADOPTED FROM [42]

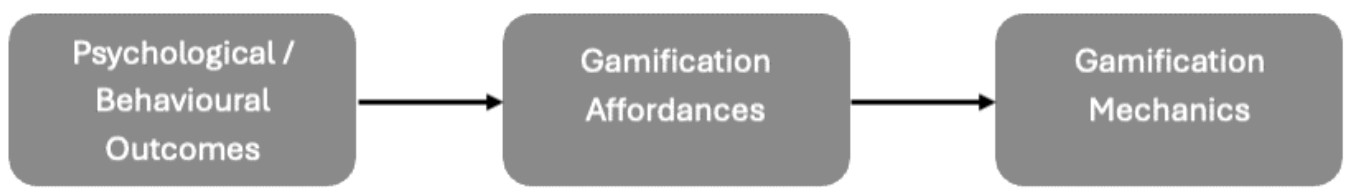

FIGURE VI. THE PGG MODEL

*a) User/player research*

We adopt a "user research comes first" stance [24], balancing tradition [23, 55, 56] with practical profiling [57-58] to infer motivations and target affordances; this is vital in an interdisciplinary [59].

*b) Defining objectives & context*

Users enter immersive spaces with intentions that map to autonomy/competence/relatedness [44]. Clarifying scenarios (social, co-creative, competitive) aligns targeted affordances with later mechanic choices.

*c) Consideration of the technologies*

Affordances and mechanics are constrained/enabled by XR hardware and platform systems (avatars, tracking, UGC, cross-platform), so design must pair user-centered intent with technological feasibility [11, 27, 28, 30, 40].

*d) Gamification Affordances*

We treat affordance mapping as dynamic and context-sensitive [26, 41-43]. When self-expression needs are high and avatar/customization tools exist, identity-construction affordances become primary and are operationalized via personalization mechanics.

*e) Gamification Design & Evaluation*

Given non-standardized affordance–mechanic mappings [41], ADGF favors process over prescription. We incorporate MDA post-hoc to assess whether chosen mechanics generate desired dynamics and aesthetic outcomes; iterative feedback refines mechanics or upstream affordances when targeted psychological outcomes are not achieved [22, 40, 44, 60].

*2) Operationalising ADGF involves five steps:*

We acknowledge this ambiguity as a current limitation of the Affordance-Driven Gamification Framework (ADGF). Existing gamification literature provides limited empirical evidence to establish definitive or standardized relationships between specific affordances and mechanics. At present, designers must rely on a small number of structured studies, such as the study [41] that proposes affordance-mechanic mappings or draw from broader frameworks that support inferential reasoning. Despite this limitation, the absence of rigid mappings introduces a valuable degree of design flexibility, encouraging practitioners to focus more deeply on the user-centered reasoning processes emphasized in earlier stages of the framework. Rather than treating mechanic selection as a fixed or formulaic outcome, the ADGF supports a process-oriented and context-sensitive design logic, one that prioritizes user needs, psychological profiles, and environmental affordances over prescriptive templates.

While gamification research has made progress in identifying common mechanics and exploring their motivational effects, the relationship between gamification mechanics and gamification affordances remains theoretically fragmented and often limited to specific domains. Most studies offer isolated or application-specific insights, lacking a widely accepted structure for aligning user motivations, perceived affordances, and designable mechanics. This theoretical gap not only complicates the design process but also increases the risk of misaligned gamification strategies.

To address this limitation, the MDA framework (Mechanics-Dynamics-Aesthetics) [40] is incorporated into the final stage of the ADGF as a reflective evaluation mechanism. Originally introduced to bridge the gap between a designer's intent and a player's experience, the MDA framework provides a structured method for analyzing how specific mechanics generate interactive system behaviors (dynamics) and produce experiential outcomes (aesthetics). This tripartite logic complements the ADGF's goal of aligning gamification design with psychological and behavioral needs.

Compared to other gamification frameworks such as Octalysis [60], Self-Determination Theory-based models [44], or Deterding's design lenses [22], the MDA model offers a layered evaluation logic that helps designers not only understand what elements they implement, but also how those elements behave in practice (dynamics) and whether they evoke the desired emotional or motivational responses (aesthetics).

While the ADGF follows a forward logic, i.e. from psychological outcomes to affordances to mechanics, the MDA framework (Mechanics-Dynamics-Aesthetics) is incorporated in the final stage as a post hoc validation tool to assess the effectiveness of this design path. In the ADGF, after gamification mechanics have been selected based on mapped affordances, the MDA framework is applied in a reflective evaluation phase. Designers analyze how the mechanics influence user-system interactions (dynamics) and whether those dynamics lead to the desired experiential outcomes (aesthetics). For example, suppose a mechanic is intended to foster self-expression. In that case, designers should observe whether users are actively customizing avatars and whether this customization contributes to emotional resonance or identity satisfaction. This process creates a feedback loop: mechanics derived through the ADGF are subjected to a comprehensive MDA analysis, beginning with mechanics, progressing to dynamics, and culminating in aesthetic outcomes such as emotional engagement or behavioral change. Notably, gamification affordances are not explicitly embedded in the MDA structure, which allows designers to indirectly assess whether the intended affordances (identified in the ADGF) have indeed been successfully activated and translated into meaningful user experiences. This step enables iterative improvement. If the targeted psychological outcomes (e.g., enhanced engagement, social bonding) are not observed, designers can revisit and adjust the mechanics or the underlying affordances to achieve the desired effects. Thus, the MDA framework acts as a systematic diagnostic tool, helping to close the design-feedback loop and ensure that gamification efforts remain aligned with user needs and contextual expectations. The incorporation of MDA reinforces both the adaptability and the theoretical coherence of the framework, supporting the development of meaningful, user-driven gamification strategies in immersive environments.

In this sense, the ADGF fosters a human-centered, iterative roadmap for gamification in the Metaverse-one in which mechanics are not rigidly defined, but emerge through cycles of contextual reasoning, behavioral validation, and dynamic refinement. Moreover, this approach aligns with the evolving nature of the Metaverse itself, which is continuously shaped by shifts in social norms, user expectations, and technological affordances. By beginning with user-specific psychological outcomes and adapting affordances and mechanics accordingly, the ADGF inherently supports customized gamification designs tailored to diverse player types, interaction contexts, and immersive environments. Operationalizing ADGF for practice involves five stages (Figure VII):

Stage 1 - User Research: identify Gen Z player types, motivations, and behavioral tendencies.

Stage 2 - Define Context & Objectives: clarify domain and target outcomes (e.g., collaboration, emotional engagement, knowledge).

Stage 3 - Technology Consideration: select XR/platform modalities and interactions.

Stage 4 - Affordance Mapping: prioritize identity, self-expression, competition, collaboration, etc., based on user insights and platform capabilities.

Stage 5 - Design & Evaluation: select mechanics and apply MDA; iterate to support flow and deliver intended behavioral/emotional outcomes.

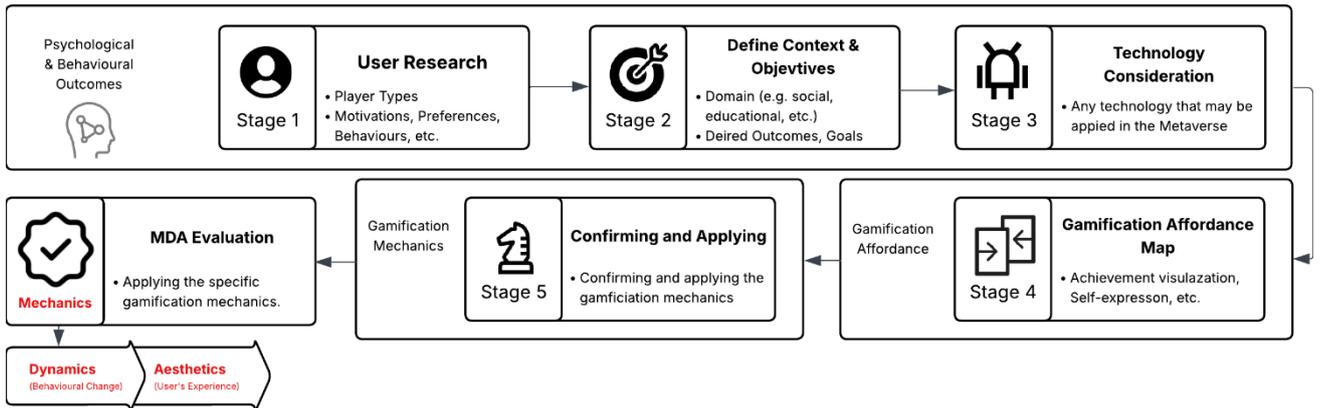

FIGURE VII. A FIVE-STEP DESIGN PROCESS FOR IMPLEMENTING THE AFFORDANCE-DRIVEN GAMIFICATION FRAMEWORK (ADGF) IN THE METAVERSE

## V. THEORETICAL CONTRIBUTIONS

With the rapid evolution of immersive technologies and the growing prominence of the Metaverse, traditional gamification frameworks originally developed for desktop, mobile, or structured domains, such as education and healthcare, are no longer sufficient to address the complexities of emerging virtual environments. These conventional models, such as Octalysis [60] and the 6D framework [23], often overlook critical contextual factors, including platform limitations, spatial interaction, and real-time co-presence. While these models offer valuable insights into player motivation and engagement, they tend to treat gamification as a top-down process, where predefined mechanics are applied regardless of technological or social constraints.

In contrast, the Affordance-Driven Gamification Framework (ADGF) proposed in this study addresses these limitations through three key theoretical contributions:

- ADGF explicitly integrates technological feasibility as a core design component. Unlike Octalysis, which centers on motivational taxonomies, or MDA, which separates mechanics from platform-level execution, ADGF embeds technological constraints such as XR modalities, input fidelity, and spatial computing into the gamification design process. This ensures that affordances and mechanics are not only psychologically resonant but also practically implementable within immersive environments.

- ADGF introduces a context-aware, perception-driven approach to affordances. Rather than selecting mechanics from a static toolkit, the framework begins with user research focused on psychological needs (e.g., autonomy, relatedness, identity). It maps these to dynamic, scenario-specific affordances such as avatar customization, co-creation, and spatial interaction. These affordances serve as bridges between user intention and technological possibility, repositioning users as active participants rather than passive recipients of gamified experiences.

- ADGF operationalizes this design philosophy into a five-step process, culminating in iterative evaluation using the Mechanics–Dynamics–Aesthetics (MDA) model. This structure enables designers to move fluidly between user insights, technological affordances, and gamified mechanics, continuously refining their approach based on behavioral outcomes. By doing so, ADGF enables a reverse-feedback loop where user behavior and psychological states inform future design iterations, particularly relevant in the co-creative, unpredictable ecosystems of the Metaverse.

Importantly, ADGF supports a paradigmatic shift in gamification design from "What do designers want users to do?" to "What do users want to do in this environment?" This inversion is especially critical in the Metaverse, where users navigate digital spaces through emergent, self-directed behaviors that defy rigid task-based structures. By anchoring mechanics in voluntary discovery, identity expression, and real-time interaction, the framework enables a form of gamification that is deeply situated, emotionally engaging, and socially emergent.

ADGF is designed to remain future-proof and extensible, adaptable to forthcoming technological paradigms such as brain-computer interfaces or fully immersive sensory platforms. Rather than prescribing fixed solutions, it offers a flexible design logic grounded in affordance theory, human-computer interaction (HCI), and user-centered design. As such, ADGF advances gamification research by establishing a theoretical bridge between technological realism and perceptual motivation, tailored to the demands of immersive social systems.

To support empirical and simulation-based validation of the Affordance-Driven Gamification Framework (ADGF), a Mixed Reality prototype titled "De-Buff Party" has been designed for deployment on the Meta Quest platform. The prototype incorporates a suite of gamified social activities tailored to Generation Z users, each systematically mapped to core affordances such as identity expression, playful competition, co-creation, and social reciprocity. These activities are structured to elicit targeted psychological outcomes including social bonding, emotional engagement, and spontaneous interaction thereby operationalizing the ADGF's five-stage design logic in an immersive environment. Although experimental trials have not yet commenced, the planned evaluation will employ pre- and post-study surveys, in-experience behavioral metrics, and post-session focus group discussions to collect both quantitative and qualitative data. This simulation-based validation pathway provides a concrete mechanism to examine the framework's affordance–mechanic mappings and reverse-feedback structure in practice. Future applications of the ADGF are envisaged across multiple domains, including co-located Mixed Reality learning spaces, avatar-driven collaborative environments, rehabilitation and well-being interventions, and socially networked Metaverse platforms. These initiatives will allow iterative refinement of the framework and establish its empirical robustness across educational, social, and affective contexts.

## VI. LIMITATIONS

Despite the systematic and theory-informed approach adopted in this review, several limitations must be acknowledged to contextualize the findings and inform future research. First, this review focused on literature published between 2020 and 2025, reflecting the recent emergence of the Metaverse as a distinct research domain. While this timeframe was necessary to capture current technological developments and user behavior trends, it may have excluded foundational gamification or XR-related studies published prior to this period. As a result, some influential early work may not have been incorporated, which could affect the historical depth of the synthesis.

Second, although the combination of PRISMA and SPIDER allowed for a structured and qualitative selection process, the final number of included articles (n = 17) is relatively small. This limitation reflects the nascent nature of Metaverse-focused gamification research and the strict inclusion criteria applied to ensure conceptual and methodological relevance. However, it may reduce the thematic breadth of the synthesis and risks over-representing educational or instructional design contexts, which dominated the selected studies. Additional research is needed to test whether the proposed framework generalizes across diverse Metaverse settings such as entertainment, marketing, open-world platforms, and decentralized social environments.

Third, while the study focused on Gen Z, not all included papers explicitly analyzed this demographic's unique characteristics. Most age-based inferences were drawn from participant samples that fit the Gen Z age range (18–30), rather than from targeted generational analyses. As a result, some insights regarding identity, motivation, and interaction preferences may not fully capture the complexity of Gen Z behaviors. Future research should incorporate explicit generational frameworks and cross-cohort comparisons (e.g., with Gen Alpha or Millennials) to validate these findings.

Fourth, the proposed Affordance-Driven Gamification Framework (ADGF) remains at a theoretical level. Although it was constructed through a rigorous thematic synthesis and operationalized into five design stages, its practical feasibility, usability, and effectiveness have not yet been empirically tested. Validation through user studies, prototype implementations, and experimental trials in immersive environments will be crucial in determining the model's real-world applicability.

Fifth, the framework assumes access to immersive technologies such as VR, AR, or MR headsets, and thus may not generalize to Metaverse experiences accessed via lower-end or mobile platforms. Future iterations of the framework should consider how gamification affordances can be adapted to varying levels of hardware sophistication and sensory input.

Finally, both "Metaverse" and "gamification" remain fluid and contested terms, lacking universally accepted definitions across disciplines. This conceptual ambiguity may have impacted both the inclusion/exclusion criteria and the synthesis process itself. Efforts were made to ensure consistent interpretation, but variation in how platforms and gamified systems are described in the literature introduces unavoidable limitations. Acknowledging these limitations provides critical context for interpreting the proposed framework. They also highlight key future directions, including empirical testing, broader domain validation, generational segmentation, and the need for conceptual clarity as the Metaverse continues to evolve.

## VII. CONCLUSION

This study aimed to investigate how gamification is conceptualized and applied in the Metaverse, with a specific focus on enhancing social interaction and sustaining engagement among Generation Z users. Using the PRISMA 2020 protocol and the SPIDER framework, a systematic literature review was conducted across multiple disciplines, including education, interaction design, and immersive technology. Three research questions guided this review: the gamification mechanics currently used in the Metaverse, the theoretical frameworks that inform them, and how these elements align with the behaviors and motivations of Gen Z.

The review findings show that although gamification is widely discussed in XR contexts, current approaches are often fragmented, domain-limited, or lack integration. Most notably, few existing frameworks directly address the social, perceptual, and identity-based engagement preferences of Gen Z users, whose interactions in immersive environments are shaped by self-expression, co-presence, and dynamic interaction styles.

To address this gap, this study introduces the Affordance-Driven Gamification Framework (ADGF), a conceptual model that connects user motivations, social affordances, and platform-specific capabilities. By focusing on identity-driven, avatar-based, and spatial interaction affordances, the ADGF provides a structure for fostering emotionally engaging and socially meaningful experiences in the Metaverse. This framework contributes theoretically by reinterpreting gamification through the lens of embodied affordance, moving beyond static mechanics to perception- and context-sensitive interaction design.

To support real-world applications, a five-step process was developed to operationalize the ADGF. This includes user research, contextual goal setting, technological consideration, affordance mapping, and MDA-informed gamification design. This stepwise model offers a practical guide for researchers and designers to implement the framework in immersive environments, bridging abstract theory with actionable design processes. The model also establishes a basis for future empirical validation through user testing and iterative refinement.

Together, the conceptual and operational contributions of this study offer a dual foundation for advancing gamified interaction in the Metaverse. The ADGF not only addresses existing theoretical gaps but also introduces a flexible, user-centered methodology for designing engaging, co-creative, and platform-aware experiences tailored to the preferences of Generation Z.

ACKNOWLEDGEMENT

N/A.

FUNDING

This research did not receive any outside funding or support. Author reports no involvement in the research by the sponsor that could have influenced the outcome of this work.

AUTHORS` CONTRIBUTIONS

All authors contributed to the drafting of the manuscript. All authors read and approved the final version of the manuscript.

CONFLICT OF INTEREST

The authors declare that they have no known competing financial interests or personal relationships that could have appeared to influence the work reported in this manuscript.

DATA AVAILABILITY

This study is a systematic literature review and relies exclusively on previously published and publicly accessible sources. No new empirical data were generated or collected for this research. All data supporting the findings of this study are available within the cited publications.

ETHICAL STATEMENT

This study is a systematic literature review and does not involve human participants, animals, or sensitive personal data. Therefore, institutional ethics approval was not required. All procedures complied with the ethical standards for research integrity and academic publishing.

DECLARATION OF AI USAGE

AI-assisted tools (e.g., ChatGPT) were used only for language refinement, grammar correction, and formatting assistance during manuscript preparation. No AI tools were used for data collection, analysis, interpretation, or for generating scientific findings, arguments, or conclusions. The authors take full responsibility for the content and integrity of the manuscript.